\documentstyle[sprocl,epsf,axodraw]{article}

\def\Journal#1#2#3#4{{#1} {\bf #2}, #3 (#4)}

\def\NPB{{\em Nucl. Phys.} B}
\def\PLB{{\em Phys. Lett.}  B}

\def\PRD{{\em Phys. Rev.} D}

\newcommand{\lsi}{\raise0.3ex\hbox{$<$\kern-0.75em\raise-1.1ex\hbox{$\sim$}}}
\newcommand{\gsi}{\raise0.3ex\hbox{$>$\kern-0.75em\raise-1.1ex\hbox{$\sim$}}}

\newcommand{\gsim}{\mathop{\gsi}}
\renewcommand{\vec}[1]{{\bf #1}}

\newcommand{\mref}[1]{(\ref{#1})}

\newcommand{\eq}[1]{Eq.\ \mref{#1}}
\newcommand{\eqs}[1]{Eqs.\ \mref{#1}}

\newcommand{\intv}[1]{\int\!\frac{d\Omega_{{\rm v}_{#1}}}{4\pi}}

\renewcommand{\(}{\left(}
\renewcommand{\)}{\right)}

\newcommand{\mmdebye}{m^2_{\rm D}}

\newcommand{\tr}{{\rm t}}

\newcommand{\vA}{\vec{A}}
\newcommand{\vE}{\vec{E}}

\newcommand{\ve}{\vec{e}}
\newcommand{\vk}{\vec{k}}
\newcommand{\vp}{\vec{p}}
\newcommand{\vq}{\vec{q}}
\newcommand{\vv}{\vec{v}}
\newcommand{\vx}{\vec{x}}
\newcommand{\mal}{\!\cdot\!}

\newcommand{\deltav}{\delta^{(S^2)}}
\newcommand{\mlabel}[1]{\label{#1}}
\newcommand{\lav}{\langle\!\langle}
\newcommand{\rav}{\rangle\!\rangle}
\newcommand{\nn}{\nonumber}

\begin{document}

\renewcommand{\theequation}{\thesection.\arabic{equation}}

\title{EFFECTIVE THEORIES FOR HOT NON-ABELIAN DYNAMICS
\footnote{ Plenary talk given at Conference on Strong and Electroweak
Matter (SEWM 98), Copenhagen, Denmark, 2-5 Dec 1998.  }}

\author{DIETRICH B\"ODEKER}

\address{Niels Bohr Institute,
Blegdamsvej 17, DK-2100 Copenhagen \O, Denmark\\E-mail: bodeker@nbi.dk}

\maketitle\abstracts{The dynamics of soft ($|\vec{p}|\sim g^2 T$)
non-Abelian gauge fields at finite temperature is
non-perturbative. The effective theory for the soft fields can be
obtained by first integrating out the momentum scale $T$, which yields
the well known hard thermal loop effective theory. Then, the latter is
used to integrate out the scale $gT$. One obtains a Boltzmann
equation, which can be solved in a leading logarithmic
approximation. The resulting effective theory for the soft fields is
described by a Langevin equation, and it is well suited for
non-perturbative lattice simulations. }

\vskip-2.75in
\rightline{NBI-HE-99-12}
\vskip2.5in

\section{Introduction} \mlabel{sec.intro}
The problem I am going to discuss is the following: How can one
calculate thermal expectation values like
\begin{eqnarray}
        C(t_1 - t_2) = \langle {\cal O}(t_1) {\cal O}(t_2)\rangle
        \mlabel{c}
\end{eqnarray}
in a non-Abelian gauge theory, when the leading order contribution is
due to spatial momenta of order $g^2 T$? The operator ${\cal O}(t)$ is
a gauge invariant function of the gauge fields $A_\mu(t,\vx)$ at real
(Minkowski-) time $t$. When I said the leading order contribution is
due to momenta of order $g^2 T$, I referred to the Fourier components
of the gauge fields entering ${\cal O}(t)$. It is not
possible to compute such a correlation function in perturbation
theory.  

This problem arises when one wants to compute the so-called hot
sphaleron rate \cite{moore}, which is the rate for electroweak baryon number
violation at very high temperatures ($T\gsim 100$GeV).  Then, the
SU(2) gauge symmetry is unbroken, and the electroweak theory is
similar to hot QCD.  Fortunately, it is a simpler because the gauge
coupling is small.

In my talk I will try to explain how such correlation functions can be
computed at leading order in the gauge coupling \cite{letter} (see also
Peter Arnold's talk~\cite{arnold}). First,
one integrates out the field modes with ``hard'' ($p\sim T$)
\footnote{For spatial vectors I use the notation
$k=|\vk|$. Four-vectors are denoted by $K^\mu = (k^0,\vk)$ and I use
the metric $K^2 = k_0^2 - k^2$.} spatial momenta (Sec.\ 3). The result is the
well known hard thermal loop effective theory. In a second step one
integrates out the modes with semi-hard ($p\sim gT$) momenta (Sec.\ 4). At
leading logarithmic order, one obtains an effective theory described
by a Langevin equation (Sec.\ 5).

\section{The classical field approximation} \mlabel{sec.classical}
Both effective theories I am going to discuss are valid for momenta
small compared to the temperature. Then, the number of field quanta in
one mode with wave vector $\vp$, given by the Bose
distribution function
\begin{eqnarray} 
	n(p)=
	\frac{1}{e^{p/T} - 1} \simeq \frac{T}{p} 
	,
\end{eqnarray}
is large.  In this case we are close to the classical field
limit. Thus, the dynamics of the low momentum modes should be governed by
classical equations of motion.

For real time problems at finite temperature, classical field theories
have a great advantage over quantum field theories. It is possible to
treat them non-perturbatively on a lattice. All one has to do is to
solve classical field equations of motion for given initial
conditions.  The solution is then inserted into the operator ${\cal
O}(t)$ to be measured. This has to be done for an ensemble of initial
configurations. The correlation function of interest is then given by the
ensemble average, where the weight is the Boltzmann factor
$\exp(-H/T)$.

The use of a classical field theory for computing the hot sphaleron
rate was suggested more than ten years ago. However, it took a long
time to understand what is the correct classical theory for the soft
modes. Originally, it was assumed that correct classical theory is
just the classical gauge theory at finite temperature. Then, Arnold,
Son and Yaffe pointed out, that the hard modes have a strong effect on
the soft dynamics \cite{asy}. Since for $p\sim T$ the Bose
distribution function is of order 1, the hard modes certainly do not
behave classically. Therefore one has to integrate them out in order
to be able to use the classical field approximation.

\begin{figure}[t]
 
 
\hspace{.8cm}
\centerline{
        \hspace{-.3cm}
        \epsfysize=6.7cm\epsffile{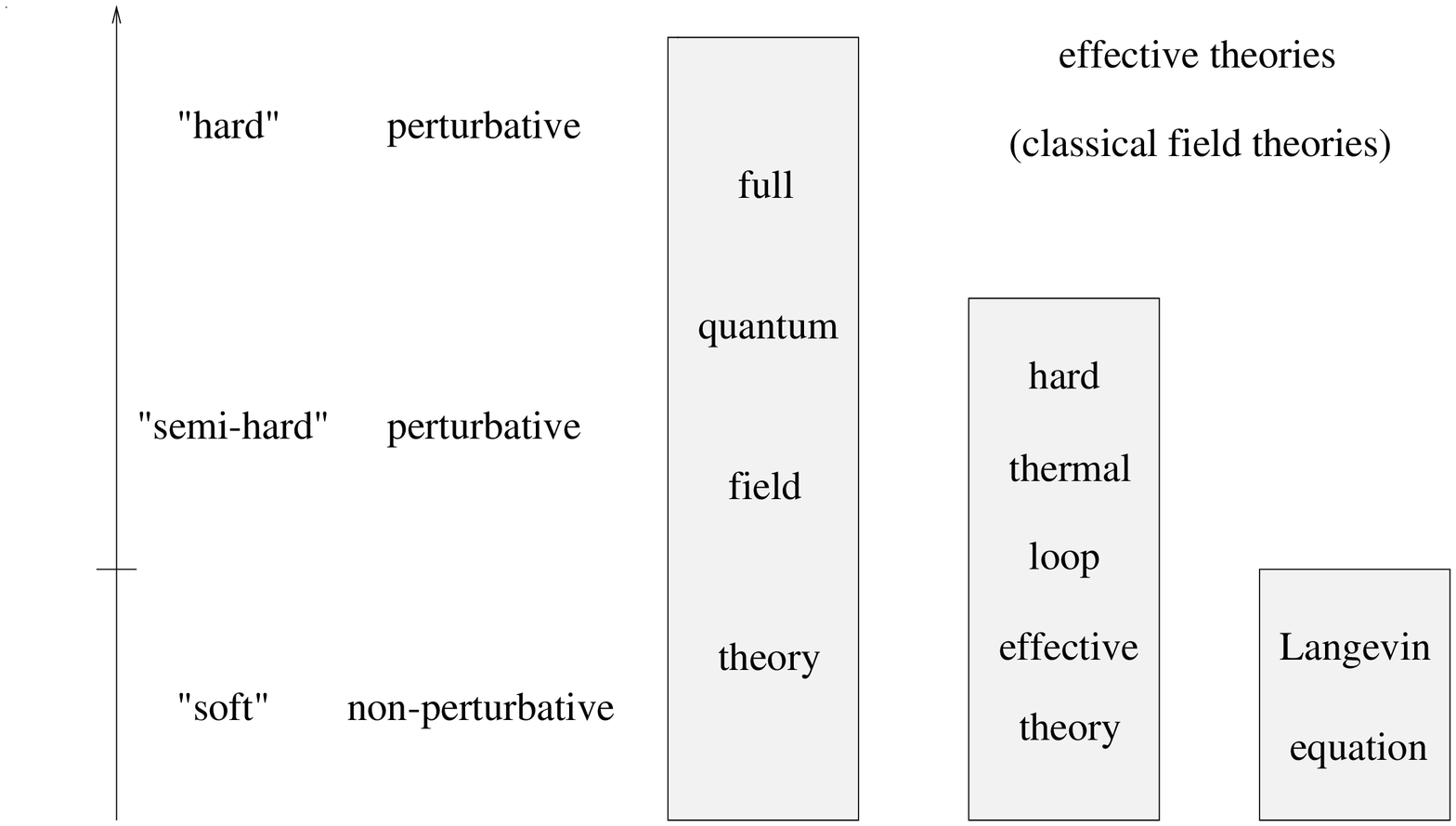}
        }

\vspace*{-3cm}
\begin{picture}(300,100)
\put(15,210){$p$}
\put(8,175.5){$T$}
\put(2,107){$g T$}
\put(6,73){$\mu$}
\put(-2,40){$g^2 T$}
\end{picture}
\caption[a]{The full quantum field theory is necessary to describe the
physics of all three relevant momentum scales, but it cannot be used
for non-perturbative calculations. The hard thermal loop effective
theory describes the physics of the scales $gT$ and $g^2 T$. Due to
Raleigh-Jeans UV divergences, it is difficult to treat
non-perturbatively. By integrating out the scale $gT$, one obtains the
Langevin equation which is free of UV problems.}  \mlabel{fig.scales}
\end{figure}

\section{Integrating out the hard modes} \mlabel{sec.t}
The hard modes constitute the bulk of degrees of freedom in the hot
plasma. Their physics is that of almost free massless particles moving
on straight lines. Even though they are weakly interacting, they have
a significant influence on the soft dynamics because they are so
numerous.

Integrating out the hard modes means that we have to calculate loop
diagrams with external momenta $\ll T$ and internal momenta of order
$T$. This generates effective propagators and vertices for the field
modes with $p\ll T$.  To leading order, we can neglect
self-interactions of the hard modes. Therefore, we can restrict
ourselves to one-loop diagrams.  The leading one-loop contribution is
due to the case that one propagator in the loop is on shell. For the
remaining propagators one can use the high energy (or eikonal)
approximation
\begin{eqnarray}
	\mlabel{eikonal.propagator}
	\frac{1}{(Q + P)^2} = \frac{1}{2 Q\cdot P+ P^2}
	\simeq \frac{1}{2 Q\cdot P}
	= \frac{1}{2 q} \frac{1}{v\cdot P}
	.
\end{eqnarray}
Here $v = (1, \vv)$, where $\vv = \vq/q$ is the 3-velocity of the
hard particles. In this way one obtains the well known hard thermal loops
\cite{pisarski}.  

Before I proceed, let me give you an argument, due to Arnold, Son and
Yaffe \cite{asy}, why hard thermal loops are relevant to the
non-perturbative dynamics of the soft modes.  For the electric, or
longitudinal modes the effect of hard thermal loops is obvious. The
longitudinal polarization tensor is of order $g^2 T^2$ and is
therefore much larger than $P^2$, when $P$ is of order $g^2 T$.  Thus
electric interactions are screened on a length scale of order
$(gT)^{-1}$.

The magnetic, or transverse modes are unscreened when $p_0$ is
zero. This leads to the well known infrared problems in hot
non-Abelian gauge theories. However, to compute unequal time
correlation functions, one has to consider non-zero real $p_0$. Then,
also the magnetic modes are screened. In order to avoid this
screening, and to develop large, non-perturbative fluctuations, the
soft modes have to move very slowly. The transverse propagator becomes
unscreened when the hard thermal loop selfenergy $\delta \Pi_\tr (P)$
becomes of order $p\sim g^2 T$.  In the small frequency limit $p_0\ll
p$, we have
\begin{eqnarray}
        \delta \Pi_\tr (P) \simeq - i\frac{\pi}{4} \mmdebye \frac{p_0}{p}
        \mlabel{deltapilimit}
	,
\end{eqnarray}
where $\mmdebye\sim (gT)^2$ is the leading order Debye mass.  From
this expression one can see that the frequency scale, at which the
propagator becomes unscreened, is $p_0\sim g^4 T$.
 
The main difference between Abelian and non-Abelian theories is that for
the former the only hard thermal loop is the polarization tensor,
while for the latter there are also hard thermal loop $n$-point
functions for all $n$. As we will see below, this has a significant
effect on the soft dynamics, it is in fact qualitatively different in
Abelian and non-Abelian theories.

The hard thermal loop effective theory is described by the effective
action
\begin{eqnarray}
	\mlabel{s.eff}
	S_{\rm eff} = S + \Gamma_{\rm HTL}
	,
\end{eqnarray}
where $ \Gamma_{\rm HTL}$ is the generating functional of the hard
thermal loop $n$-point functions. It is gauge
invariant and non-local.  The non-locality is due to the eikonal
propagators like in \eq{eikonal.propagator}.

As I discussed above, this effective theory is a classical field
theory, since both $gT$ and $g^2 T$ are small compared to
$T$. Therefore, it is described by the classical
equation of motion $\delta S_{\rm eff}/\delta A_\mu=0$. As the
effective action itself, this equation of motion is non-local, which
makes it difficult to deal with.

\begin{figure}[t]
 
 
\hspace{.8cm}
\epsfysize=5cm
\centerline{
        \hspace{-.3cm}
        \epsfysize=6.7cm\epsffile{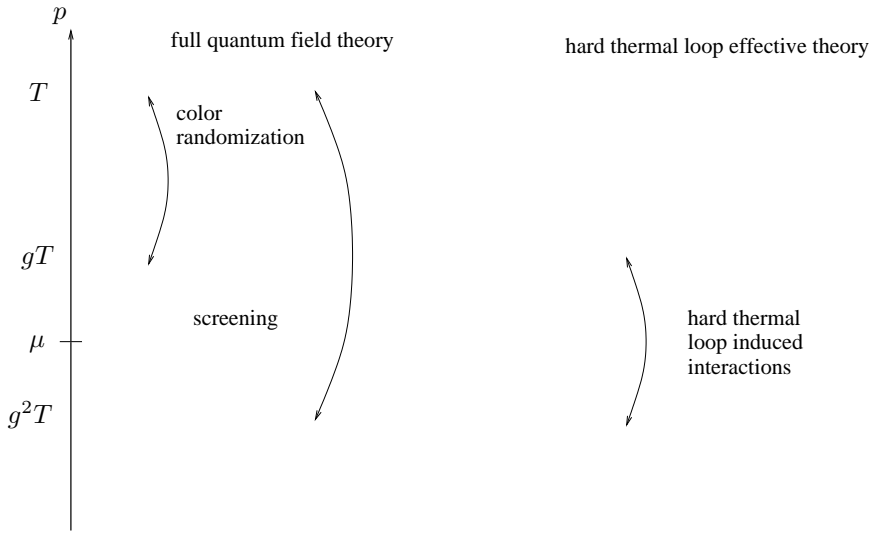}
        }

\vspace*{-3cm}
\begin{picture}(300,100)
\put(15,210){$p$}
\put(6,179){$T$}
\put(3,117){$g T$}
\put(6,86){$\mu$}
\put(-2,57){$g^2 T$}
\end{picture}

\caption[a]{The relevant interactions between the different field
modes. In the full theory, only the hard--semi-hard and hard--soft
interactions contribute at leading order. Therefore, the relevant
interactions within the hard thermal loop effective theory are due to
hard thermal loop vertices.} \mlabel{fig.interactions}
\end{figure}

Fortunately, there is a local formulation of the equations of motion
due to Blaizot and Iancu \cite{blaizot93} and due to Nair
\cite{Nair:local}. It is the non-Abelian generalization of the
linearized Vlasov equations for relativistic QED plasma. In addition
to gauge fields, these equations contains a field $W(x,\vv)$ which
lives in the adjoint representation. It describes the fluctuation of
the distribution of hard particles with 3-velocity $\vv$
(cf. \eq{eikonal.propagator}) around thermal equilibrium. The equation
of motion for the gauge fields is
\begin{eqnarray}
        \mlabel{maxwell}
        [D_\mu, F^{\mu\nu}(x)] &=&  \mmdebye \int\frac{d\Omega_\vec{v}}{4\pi}
        v^\nu  W(x,\vv)
	.
\end{eqnarray}
The rhs is the current due to the hard particles. The equation for 
$W$ reads
\begin{eqnarray}
        \mlabel{vlasov}
        [v\cdot D, W(x,\vv)] &=& \vec{v}\cdot\vec{E} (x)
	,
\end{eqnarray}
where $\vE$ is the (color-) electric field.

One may wonder, why one does not stop at this point and uses
\eqs{maxwell}, \mref{vlasov} for a lattice calculation. The reason is,
that they suffer from Raleigh-Jeans UV divergences \cite{bms}. So far,
no method has been found which cures these divergences in real time
correlation functions.  Fortunately, perturbation theory still works
for momenta of order $gT$. Therefore, one can integrate out this scale
and, at leading log accuracy, the resulting effective theory is free
of UV problems.

\section{Integrating out the semi-hard modes} \mlabel{sec.gt}
The fields in the kinetic equations \mref{maxwell}, \mref{vlasov}
contain Fourier components with momenta of order $gT$ and $g^2 T$. Now
we split these fields into long- and short wavelength components.  We
introduce a separation scale $\mu$ such that
\begin{eqnarray}
        g^2 T\ll \mu \ll gT. 
\end{eqnarray}
The fields $A$, $\vec{E}$ and $W$ are decomposed into soft and a
semi-hard modes \footnote{For notational simplicity, no new symbols
are introduced for the soft modes. From now on $A$, $\vec{E}$ and $W$
will always refer to the soft fields only.},
\begin{eqnarray}
   A \to  A + a ,\quad
   \vec{E} \to  \vec{E} + \vec{e} ,\quad
   W \to W + w 
   \mlabel{separation}
	.
\end{eqnarray}
The soft fields $A$, $\vec{E}$ and $W$ contain
the spatial Fourier components with $p<\mu$, while the semi-hard fields
$a$, $\ve$ and $w$ consist of those with $k>\mu$.  Both $W$ and $w$
describe deviation of the distribution of the hard particles from
thermal equilibrium. $W$ ($w$) is the slowly (rapidly) varying piece
of this distribution varying on length scale greater (less) than
$1/\mu$.

Integrating out the scale $gT$ means that we eliminate the semi-hard
fields from the equations of motion for the soft ones. Then we will obtain
equations of motion for the soft fields only \footnote{For details, see
\cite{letter,ladder,kinetic}.}. 

After the split one obtains a set of coupled equations of motion for
the soft and for the semi-hard fields. The low momentum
part of \eq{vlasov} becomes
\begin{eqnarray}
        \mlabel{vlasovsoft}
        [v \cdot D, W(x,\vv)]   = \vec{v}\cdot\vec{E} (x)
        + \xi(x,\vv)
	,
\end{eqnarray}
where
\begin{eqnarray}
        \mlabel{F.0.5.1.5}
        \xi (x,\vv) = i g [v\mal a(x), w(x,\vv)]_{\rm soft}
	.
\end{eqnarray}
The subscript ``soft'' indicates that only spatial Fourier components
with $p<\mu$ are included.

I said that the semi-hard fields are perturbative. Here we are only
interested in leading order results. Then one might expect, that one
can approximate $\xi\simeq \xi_0$, where 
\begin{eqnarray}
        \mlabel{xi0}
        \xi_0 (x,\vv) = i g [v\mal a_0(x), w_0(x,\vv)]_{\rm soft}
	,
\end{eqnarray}
and $a_0$, $w_0$ are the solutions to the linearized
kinetic equations. This is not quite correct, which
will become clear in the moment. But for the sake of simplicity, let
us assume that $\xi\simeq \xi_0$ is a good approximation, and consider
the effect of $\xi_0$ in \eq{vlasovsoft}. 

In order to compute correlation functions, one has to solve the
equations of motion for the soft fields in the presence of
$\xi_0$. Since these equations are non-linear, the solution
will contain many factors of $\xi_0$. Now one has to perform the
thermal average over initial conditions. One encounters 
expectation values like\begin{eqnarray}
	\langle \xi_0(x_1,\vv_1)\cdots
        \xi_0 (x_{n},\vv_{n}) \rangle  
	\mlabel{xi0.npoint}
	.
\end{eqnarray}
The typical separation of the points $\vx_1,\ldots,\vx_n$ is of order
$(g^2 T)^{-1}$. In contrast, the fields $a_0$ and $w_0$ are correlated
over a much smaller length scale of order $(g T)^{-1}$. Therefore,
\mref{xi0.npoint} factorizes into a product
\begin{eqnarray}
	\langle \xi_0(x_1,\vv_1)\cdots
        \xi_0 (x_{n},\vv_{n}) \rangle
	\simeq
	\langle \xi_0(x_1,\vv_1)\rangle  \cdots
        \langle \xi_0 (x_{n},\vv_{n}) \rangle  
	\mlabel{xi0.disconnected}
	,
\end{eqnarray}
while connected parts are suppressed by some powers of the coupling
constant. Now we will see why it is not sufficient to use the
approximation $\xi\simeq\xi_0$.  The rhs of \eq{xi0.disconnected} is zero!
$\langle \xi_0^a\rangle$ contains the
expectation value
\begin{eqnarray}
        \langle a_0^b w_0^c \rangle \propto \delta^{bc}
	\mlabel{a0w0}
	,
\end{eqnarray}
which is contracted the anti-symmetric structure constant $f^{abc}$.
Therefore, in order to obtain the leading non-vanishing contribution
for \mref{xi0.npoint}, one has to take into account connected 2-point
functions of $\xi_0$.  In other words, $\xi_0$ acts like a Gaussian
noise.

Since the leading order contribution due to $\xi_0$ vanishes, one also
has to take into account the first ``sub-leading'' term in $\xi$
itself, which will be denoted by $\xi_1$. It is linear in the soft
fields, and, like $\xi_0$, it is bilinear in $a_0$ and $w_0$.  However,
in this case the thermal average \mref{a0w0} gives a non-zero
contribution. Thus, at leading order, one can approximate
\begin{eqnarray}
	\xi (x,\vv) \simeq \xi_0 (x,\vv) + \lav \xi_1(x,v) \rav
	 \mlabel{xi.right}
	,
\end{eqnarray}
where $\lav \cdots \rav$ denotes the average over initial conditions
for $a_0$ and $w_0$. 

Evaluating the 2-point function of $\xi_0$, one encounters a
contribution which is logarithmically sensitive to the separation
scale $\mu$. Keeping only this piece, one finds
\begin{eqnarray}
        \langle 
    \xi_0^{a}(x_1,\vv)
    \xi_0^{b}(x_{2},\vv') 
    \rangle 
    &=&
    -\frac{2 N g^2 T^2}{\mmdebye}
        \log\(\frac{g T}{\mu}\)
         I(\vv,\vv')
\nn\\ &&
        \delta^{ab} \delta^4(x_1 - x_2)
        \mlabel{xi0correlator4}
	,
\end{eqnarray}
with
\begin{eqnarray}
  \mlabel{k}
  I(\vv,\vv') = -\deltav(\vv - \vv') 
        + \frac{1}{\pi^2}
        \frac{(\vv\cdot\vv')^2}{\sqrt{1 - (\vv\cdot\vv')^2}} 
	.
\end{eqnarray}
Here, $\deltav$ is the delta function on the two dimensional unit
sphere,
\begin{eqnarray}
        \int d\Omega_{\vv'} f(\vv') \deltav(\vv - \vv') = f(\vv) 
	.
\end{eqnarray}
The term $\lav\xi_1\rav$ contains a piece which has the same
$\mu$-dependence as \mref{xi0correlator4}.  Inserting the result into
\eq{vlasovsoft}, one finds
\begin{eqnarray}
        [v \cdot D, W(x,\vv)] &=& \vec{v}\cdot\vec{E} (x)+ \xi_0(x,\vv)
        \nn\\ &&
        \hspace{-2cm} {}  + N g^2 T  \log\(\frac{g T}{\mu}\)
 \intv{1}   I(\vv,\vv_1)      
 W(x,\vv_1) 
        \mlabel{boltzmann}
	.
\end{eqnarray}

Eq.~\mref{boltzmann} is a Boltzmann equation for the soft fluctuations
of the particle distribution $W(x,\vv)$. The rhs contains a collision
term which is due to the interactions with the semi-hard fields.  The
collision term is accompanied by the Gaussian white noise $\xi_0$,
which is due to the thermal fluctuation of initial conditions of the
fields with $k>\mu$.

For a QED plasma, there is no collision term at this order in the
coupling constant. In this case the size of the collision term is
determined by the transport cross section which corresponds to a mean
free path of order order $(e^4 T)^{-1}$.  For a non-Abelian plasma the
relevant mean free path \cite{gyulassy} is of order $(g^2 T \log(1/g)
)^{-1}$. It is determined by the total cross section which is
dominated by small angle scattering: Even a scattering process which
hardly changes the momentum of a hard particle can change its color
charge which is what is seen by the soft gauge fields.

\section{Solving the Boltzmann equation}
\mlabel{sec.solving}
I will now argue, that, at leading logarithmic order, the lhs of
\eq{boltzmann} can be neglected. The argument goes as follows
\footnote{For details, see \cite{kinetic}.}: The only spatial momentum
scales which are left in the problem are $\mu$ and $g^2 T$. The field
modes we are ultimately interested in, are the ones which have only
momenta of order $g^2T$.  The cutoff dependence on the rhs must drop
out after solving the equations of motion for the fields with spatial
momenta smaller than $\mu$.  Thus, after the $\mu$-dependence has
cancelled, the logarithm must turn into $\log(gT/(g^2 T)) =
\log(1/g)$.

\begin{figure}[t]
 
 
\hspace{.8cm}
\epsfysize=6.5cm
\centerline{
        \hspace{-.3cm}
        \epsfysize=5cm\epsffile{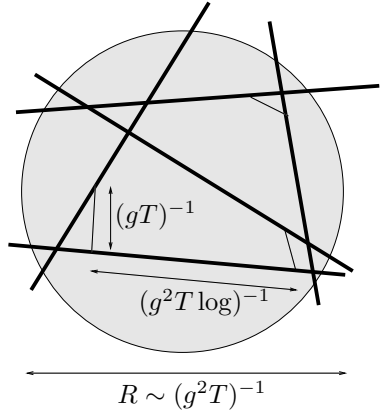}
        }

\vspace*{-3cm}
\begin{picture}(300,100)
%
\put(139,73){$(g T)^{-1}$}
\put(148,40){$(g^2 T\log)^{-1}$}
\put(140,5){$R\sim (g^2 T)^{-1}$}
\end{picture}

\caption[a]{The non-perturbative dynamics is due to extended field
configurations (shaded) with a typical size $R$ of order $(g^2
T)^{-1}$. Since the soft fields are changing in time, they are
associated with a long wavelength electric field.  The hard modes
behave like almost free particles moving on light-like trajectories
(thick lines).  They absorb energy from the long wavelength electric
field which leads to the (Landau-) damping of the soft dynamics. The
semi-hard field modes (thin lines) are responsible for color changing
small angle scattering of the hard particles. In the leading log
approximation (Sect.~\ref{sec.solving}), the typical distance between
these scattering events is small compared to $R$, so that the damping
becomes local.}  \mlabel{fig.scattering}
\end{figure}
Then the Boltzmann equation can be solved in logarithmic accuracy,
i.e., neglecting terms which are suppressed by inverse powers of
$\log(1/g)$.  The collision term on the rhs is logarithmically
enhanced over the flow term on the lhs, and one can neglect the lhs
altogether. In other words, the kinematic of the hard particles does
not play a role in this approximation (for a physical picture, see
Fig.~\ref{fig.scattering}). Multiplying \mref{boltzmann} with $v^i$
and integrating over $\vv$, one obtains
\begin{eqnarray}
        \frac13 E^i(x) 
       {}+ \xi_0^i(x) 
        -  \frac{1}{4\pi}  N g^2 T\log(1/g) W^i(x) = 0
        \mlabel{wi.2}
	,
\end{eqnarray}
where
\begin{eqnarray}
        W^{i} (x) &=& \intv{} v^{i} W(x,v)
        ,
\\
        \xi_0^{i} (x) &=& \intv{} v^{i} 
        \xi_0(x,v) 
	.
\end{eqnarray} 
Solving \mref{wi.2} for $W^i$, and inserting the result into Maxwell's
equation for the soft fields, the latter becomes (in $A_0 = 0$ gauge)
\begin{eqnarray}
        \ddot{ A}^i +  [D_j,F^{j i}(x)] = 
        - \gamma \dot{A}^i(x) + \zeta^i(x)
        \mlabel{langevin1}
	,
\end{eqnarray}
where the damping coefficient (or color conductivity \cite{arnold})
is given by
\begin{eqnarray}
        \gamma = \frac{4\pi\mmdebye}{3 Ng^2 T\log(1/g)}
	.
\end{eqnarray}
The noise $\zeta^i$ is proportional to $\xi_0^i$, 
\begin{eqnarray}
        \zeta^i(x) = \frac{4\pi\mmdebye}{Ng^2T\log(1/g)}
         \xi_0^i(x)
	.
\end{eqnarray}
Its 2-point function can be obtained from \eq{xi0correlator4},
\begin{eqnarray}
        \langle 
    \zeta^{ia}(x_1)
    \zeta^{jb}(x_{2}) 
    \rangle 
        =
    2 T \gamma
        \delta^{ij } \delta^{ab}  \delta^4(x_1 - x_2)
        \mlabel{langevin2}.
\end{eqnarray}

\section{The Langevin equation}
\mlabel{sec.langevin}
The equation of motion \mref{langevin1} describes an over-damped
system.  To see this, let us estimate \footnote{This estimate does not
rely on perturbation theory. For the soft modes both terms in the
covariant derivative $\partial_i - g A_i$ are of the same order
because of $A_i \sim gT$.} the second term on the lhs. It contains two
covariant space derivatives of the gauge fields. Each derivative is of
order $g^2 T$ . Thus, this term can be estimated as
\begin{eqnarray}
	[D_j,F^{j i}] \sim  g^4 T^2 \vA
	\mlabel{lhs}.
\end{eqnarray}
For the damping term on the rhs we have
\begin{eqnarray}
	\gamma \dot{\vA} \sim \frac{T}{\log(1/g)} \frac{ \vA}{t}
	\mlabel{rhs}.
\end{eqnarray}
where $t$ is the characteristic time scale of the soft fields.
Comparing \mref{lhs} and \mref{rhs}, we find
\begin{eqnarray}
  t^{-1}\sim g^4 \log(1/g)T 
	\mlabel{timescale}.
\end{eqnarray}
Therefore, the second time derivative in \mref{langevin1}  is
negligible, and the dynamics of the soft modes is correctly
described by the Langevin equation 
\begin{eqnarray}
        [D_j,F^{j i}(x)] = 
        -\gamma \dot{A}^i(x) + \zeta^i(x)  
        \mlabel{langevintag}
	.
\end{eqnarray}
The innocent looking approximation of dropping the term $\ddot{\vA}$ has
an important effect. The effective theory described by \eqs{langevintag},
\mref{langevin2}
is no longer sensitive to an UV cutoff \cite{asy2}. Therefore it has a 
continuum limit when used in lattice simulations. 
\section{The hot sphaleron rate} 
\mlabel{sec.sphaleron}
Now that we know that the soft non-perturbative dynamics of the
gauge fields is correctly described by \eq{langevintag}, and that
there is no dependence on the UV cutoff, it is straightforward to 
estimate the parametric form of the hot sphaleron rate. There is
only one length scale $R\sim(g^2 T)^{-1}$, and only one time scale 
$t\sim (g^4 \log(1/g)T)^{-1}$ left
in the problem, so that we can estimate
\begin{eqnarray}
	\Gamma \sim \frac{1}{t R^3} \sim g^{10} \log(1/g) T^4
	.
\end{eqnarray}
Therefore, at leading logarithmic order, the hot sphaleron rate has
the form
\begin{eqnarray}
  \Gamma = \kappa g^{10}\log(1/g)T^{4}
	\mlabel{rate.2}
	,
\end{eqnarray}
where $\kappa$ is a non-perturbative coefficient which does not depend
on the gauge coupling and which has been determined by solving 
\mref{langevintag} on the lattice \cite{moore}.

\section{Summary} \mlabel{sec.summary}
We have obtained an effective theory for the non-perturbative dynamics
of the soft field modes by integrating out the hard ($p\sim T$) and
semi-hard modes ($p\sim gT$) in perturbation theory.  This effective
theory is described by the Langevin equation \mref{langevintag}.

Furthermore, we have determined the parametric form of the hot
electroweak baryon number violation rate at leading order. It contains a
non-perturbative numerical coefficient which can be evaluated using
\eq{langevintag}.

\vspace{1cm}
{\bf Acknowledgment} This work was supported in part by the TMR network
``Finite temperature phase transitions in particle physics'', EU
contract no. ERBFMRXCT97-0122.

\end{document}